\documentclass[]{aa}

\usepackage{graphicx}
\usepackage{txfonts}
\usepackage{chemformula} 
\usepackage{threeparttable}

\graphicspath{{./}{figures/}}

\begin{document}

\title{\ch{CF+} excitation in the interstellar medium}

\author{
                 Benjamin Desrousseaux\inst{1}
\and Fran\c cois Lique\inst{1}
\and Javier R. Goicoechea\inst{2}
\and Ernesto Quintas-S\'anchez\inst{3}
\and Richard Dawes\inst{3}
}

\institute{
             LOMC - UMR 6294, CNRS-Universit\'e du Havre, 25 rue Philippe Lebon, BP 1123 - 76063 Le Havre cedex, France
\and Instituto de F\'{\i}sica Fundamental (CSIC). Calle Serrano 121-123, 28006, Madrid, Spain.
\and Department of Chemistry, Missouri University of Science and Technology, Rolla, Missouri 65409, United States
}

\abstract{
        The detection of \ch{CF+} in interstellar clouds
        potentially allows astronomers to infer the elemental fluorine abundance and the ionization fraction in ultraviolet-illuminated molecular gas.
        Because local thermodynamic equilibrium (LTE) conditions are hardly 
        fulfilled in the interstellar medium (ISM), the accurate determination of the \ch{CF+} abundance requires one to model its \mbox{non-LTE} excitation via both radiative and collisional processes.
        Here, we report quantum calculations of rate coefficients for the rotational excitation of \ch{CF+} in collisions with para- and ortho-\ch{H2} (for temperatures up to 150~K).
        As an application, we present \mbox{non-LTE} excitation models that reveal population inversion in physical conditions typical of ISM photodissociation regions (PDRs). 
        We successfully applied these models to fit the \ch{CF+} emission lines previously observed toward the Orion Bar and Horsehead PDRs.
        The radiative transfer models achieved with these new rate coefficients allow the use of \ch{CF+} as a powerful probe to study molecular clouds exposed to strong stellar radiation fields.
}

\maketitle

\section{Introduction}
Although fluorine is one of the most reactive species in the interstellar medium (ISM; \cite{neufeld:2005a, neufeld:2009}), the chemistry of the most abundant F-bearing molecules, HF and \ch{CF+}, can be accurately described by a few chemical reactions that depend on the amount of F atoms, H$_2$ molecules, and C$^+$ ions \citep{neufeld:2005a}.
\ch{CF+} is produced in ISM regions bathed by stellar ultraviolet (UV) photons able to ionize carbon atoms in diffuse interstellar clouds or at the illuminated surfaces of dense molecular clouds (so-called photodissociation regions; PDRs).
The simplicity of the fluorine chemical network results in a great sensitivity of astrophysical models to molecular data (reactive rate coefficients).

Interstellar \ch{CF+}  was first detected in the Orion Bar PDR by \citet{neufeld:2006}.
Since then, this molecular ion has been widely observed in other PDRs and diffuse clouds of the Milky Way and beyond \citep{neufeld:2006,kalenskiiSpectralScanStarforming2010,kalenskiiSpectralSurveyStarforming2010,guzmanIRAM30mLineSurvey2012,lisztHCOCCCF2014a,lisztWidespreadGalacticCF2015,mullerDetectionExtragalacticCF2016,nagy:2013,nagy:2017}.
In these  environments, the \ch{CF+} abundance is a powerful proxy for \ch{C+} (closely related to the ionization fraction) and for F atoms, which are both difficult to observe \citep{guzmanIRAM30mLineSurvey2012}.
Because local thermodynamic equilibrium (LTE) conditions are hardly fulfilled in \mbox{the ISM} \citep{roueffMolecularExcitationInterstellar2013a}, the competition between radiative and collisional processes has to be taken into account in the molecular line modeling in order to derive an accurate \ch{CF+} abundance and gas physical conditions.
Collisional data for collisions between \ch{CF+} and the most abundant species in the ISM (usually atomic and molecular hydrogen) are then essential.
Collisional rate coefficients with \ch{He}, as a template for \ch{H2}, were first computed by \citet{ajili:2013} and recently updated by \citet{denis-alpizar:2018}.
However, it is well established \citep{roueffMolecularExcitationInterstellar2013a} that \ch{He} data are a bad surrogate for \ch{H2} in case of collisions with an ion.

\citet{denis-alpizar:2019} provided the first proper study of the \ch{CF+}--\ch{H2} system.
They treated the rotational relaxation of \ch{CF+} by para-\ch{H2} using a reduced-dimensional potential energy surface (PES) averaged over \ch{H2} rotation.
The use of an averaged PES, however, only allows one to consider collisions of \ch{CF+} with a spherical \ch{H2}, neglecting anisotropy effects due to the \ch{H2} rotation.

\ch{CF+} is mostly detected in warm molecular gas, where the \ch{H2} ortho-to-para ratio is found to be large, therefore obtaining rate coefficients for collisional excitation with ortho-\ch{H2} is crucial and cannot be achieved considering a spherical structure-less \ch{H2}.

In order to overcome this limitation, \citet{desrousseaux:2019a} recently presented a new highly accurate four-dimensional (4D) PES, by which the calculation of rate coefficients for collisional excitation of \ch{CF+} by both para- and ortho-\ch{H2} is made possible.
In that work, preliminary cross-sections at low collisional energy (up to 150~cm$^{-1}$) were presented for both para- and ortho-\ch{H2} collisions.
In this paper, we describe further results using this recently computed \ch{CF+}--\ch{H2} 4D PES~ \citep{desrousseaux:2019a} to carry out scattering calculations of rate coefficients for the collisional excitation of \ch{CF+} by \ch{H2}.
We also used the new collisional data to simulate the excitation of \ch{CF+} in the ISM.
We demonstrate that \ch{CF+} exhibits population inversion at typical physical conditions of molecular clouds, and we show that \ch{CF+} is a powerful tracer of the gas characteristics in diffuse interstellar clouds and illuminated surfaces of dense molecular clouds.

The \ch{CF+}–\ch{H2} PES was computed employing electronic structure data from coupled-cluster theory extrapolated to the complete basis set (CBS) limit and constructed using the recently released software-package \textsc{autosurf} \citep{quintas-sanchez:2019}.
On this PES, a global minimum of $-1230.2$~cm$^{-1}$ is found for a planar structure with the two monomers almost parallel (see \citet{desrousseaux:2019a} for more details). 
This large energy well-depth, typical for ion-molecule collisional systems, makes employing state-of-the-art methods to compute the cross-sections very challenging, and, to the best of our knowledge, only four molecular ions (\ch{CN-} \citep{klos:2011}, \ch{HCO+} \citep{masso:2014}, \ch{C6H-} \citep{walker:2016a}, and \ch{SH+} \citep{2019JChPh.150h4308D, 2019MNRAS.487.3427D}) among the 38 detected in the ISM have been studied in collision with nonspherical \ch{H2}.

\section{Scattering calculations}

The \textsc{molscat} nonreactive scattering code (Hutson \& Green 1994) was used to carry out close-coupling calculations of the rotational (de)excitation cross-sections for collisions between \ch{CF+} and both para- and ortho-\ch{H2}.
The cross-sections were computed using the quantum time-independent close-coupling (CC) approach and the hybrid log-derivative/Airy propagator implemented in the \textsc{molscat} code. 
In the following, molecule-related parameters are labeled with subscripts 1 and 2, referring to the \ch{CF+} and \ch{H2} molecules, respectively.

The expansion over angular functions of the potential was performed as described by \citet{green:1975}:
\begin{equation}
V(R, \theta_1, \theta_2, \varphi) = \sum_{l_1, l_2, l} v_{l_1, l_2, l}(R) A_{l_1, l_2, l}(\theta_1, \theta_2, \varphi)
,\end{equation}
where $A_{l_1,l_2,l}(\theta_1,\theta_2,\varphi)$ is constructed from coupled spherical functions $Y_{l_i,m_i}(\theta_i,\varphi)$ and the rotational angular momenta of \ch{CF+} and \ch{H2}. 
The potential was expanded including $0 \le l_1 \le 24$ for the \ch{CF+} molecule,  and $0 \le l_2 \le 6$ for the \ch{H2} molecule.

The two molecules were considered as rigid rotors. 
The molecular constants of \ch{CF+} were set at the value given by \citet{cazzoli:2010}: $B_e=1.720912$ cm$^{-1}$, $\alpha_e=0.0189$ cm$^{-1}$, and $D_e=63 \times 10^{-6}$ cm$^{-1}$.
The molecular constants of \ch{H2} were set as \citep{herzberg:1979} $B_e=60.853$ cm$^{-1}$, $\alpha_e=3.062$ cm$^{-1}$, and $D_e=4.71 \times 10^{-2}$ cm$^{-1}$.

Cross-sections were obtained for the first 22 rotational levels of \ch{CF+} ($0 \le j_1 \le 21$).
As already discussed in \citet{desrousseaux:2019a}, convergence of these cross-sections was ensured by including the 33 lowest rotational levels of \ch{CF+} and only the lowest rotational levels of para-\ch{H2} ($j_2=0$) and ortho-\ch{H2} ($j_2=1$) in the basis set.

At each collisional energy, the maximum value of the total angular momentum $J_{\text{tot}}$ was automatically determined by the \textsc{molscat} code in order to converge cross-sections to better than $1\times 10^{-4}$ \AA$^2$, going up to $J_{\text{tot}}=120$ at the highest energies.

The determination of the thermal rate coefficients was achieved by averaging the cross-sections $\sigma_{\alpha\rightarrow \beta}$ over the collisional energy ($E_c$):
\begin{equation}
k_{\alpha\rightarrow \beta}(T)=\left( \frac{8}{\pi \mu k_B^3 T^3 } \right)^{\frac{1}{2}} \times
\int_{0}^{\infty} \sigma_{\alpha\rightarrow \beta}\ E_c e^{-\frac{E_c}{k_{\text{B}} T}}dE_c,
\end{equation}
where $\mu$ is the reduced mass of the colliding system and $k_B$ is the Boltzmann's 
constant.
$\alpha$ and $\beta,$ respectively, designate the initial and final rotational states of the \ch{CF+} molecule.
Collisional energies up to 1500 cm$^{-1}$ were explored in order to allow the determination of the rate coefficients for collisions between \ch{CF+} and both para- and ortho-\ch{H2} up to 150~K for the first 22 rotational levels of \ch{CF+}.

\begin{figure}
        \centering
        \includegraphics[width=\linewidth]{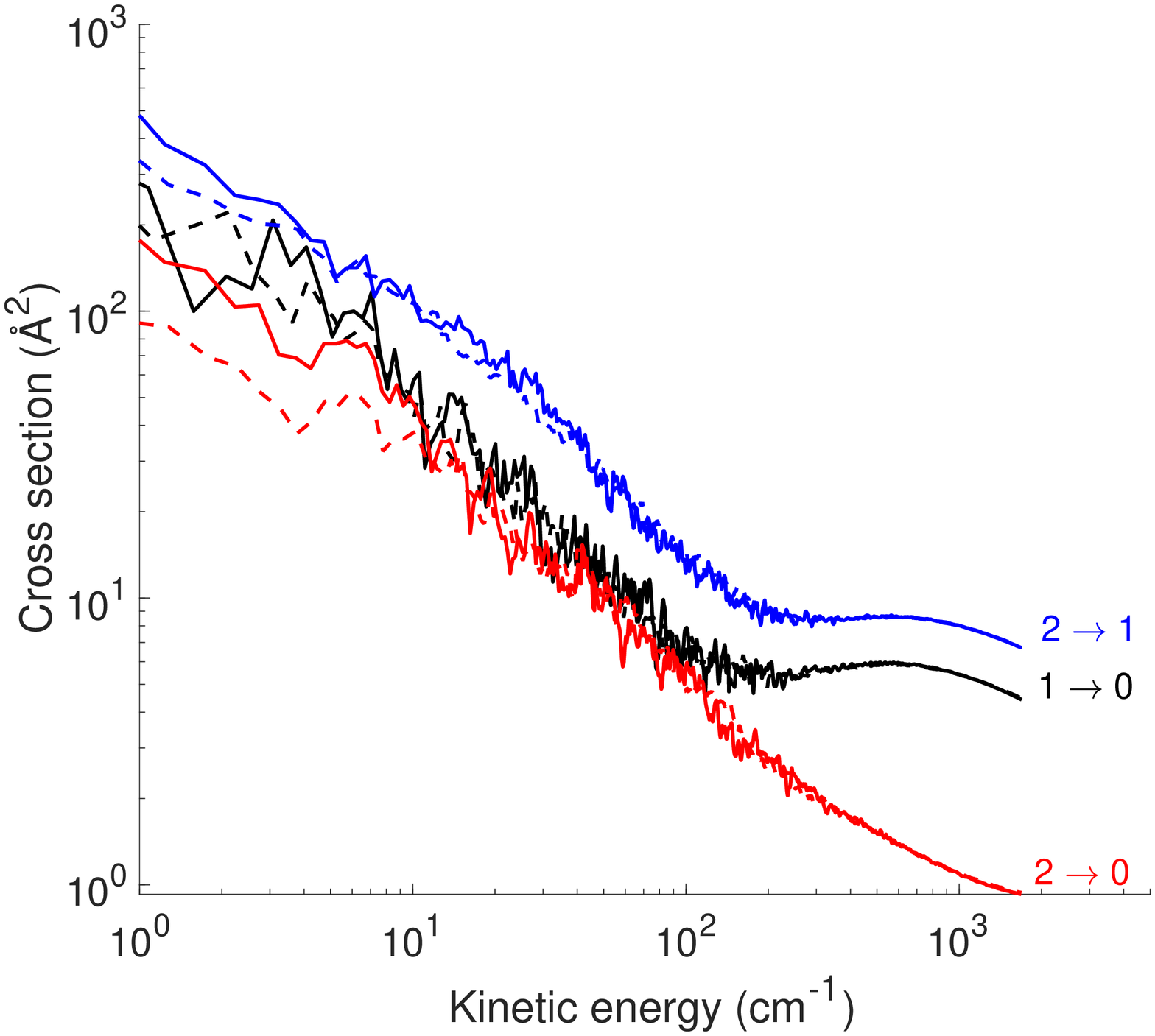}
        \includegraphics[width=\linewidth]{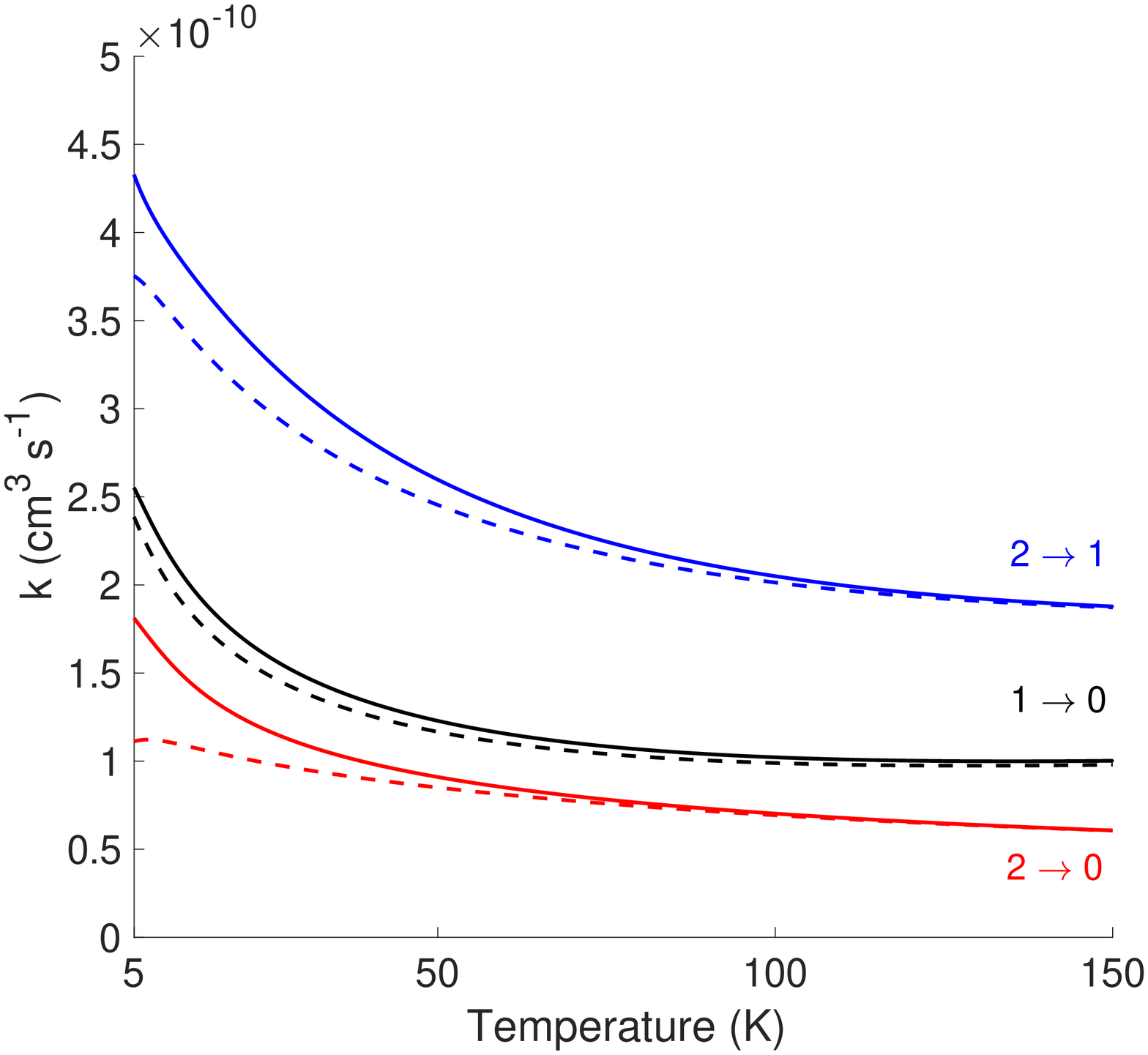}
        \caption{Cross-sections (upper panel) and rate coefficients (lower panel) for the collisional excitation of \ch{CF+} by para-\ch{H2}($j_2=0$) (solid lines) and ortho-\ch{H2}($j_2=1$) (dashed lines) for selected rotational transitions $j_1 \rightarrow j_1'$.}
        \label{xsra}
\end{figure}

Cross-sections and corresponding rate coefficients for some selected transitions are displayed in Figure \ref{xsra}.
The cross-sections (upper panel) present the same behavior for both para- and ortho-\ch{H2}.
At low energy (up to a few hundreds cm$^{-1}$), the cross-sections decrease with increasing energy, following an almost linear dependence as a function of the energy logarithm.
This behavior is typical and expected, as predicted from Langevin theory for ion--molecule collisions.
The cross-sections also exhibit many resonances in this energy region.
This can be explained by the creation of quasi-bound states within the deep van der Waals well of the \ch{CF+}--\ch{H2} complex before its dissociation, as already discussed by \citet{denis-alpizar:2020}.
At higher energies, it is generally observed that the cross sections are slowly decreasing with increasing energy, beyond a slight initial increase for transitions with low $\Delta j_1$.
The magnitude of the cross-sections seems to be approximately the same for collisions with either para- or ortho-\ch{H2}, except at very low collisional energies ($< 10$ cm$^{-1}$), where differences of up to a factor of 2 can be seen for large $\Delta j_1$ transitions.

The corresponding rate coefficients (lower panel) show a relatively flat temperature dependence, as predicted by the Langevin theory.
The magnitude of the rate coefficients decreases with increasing $\Delta j_1$, with propensity rules in favor of transitions with $\Delta j_1=1$.
Not surprisingly, given the similar cross-sections, it is also observed that rate coefficients for collisions with para- and ortho-\ch{H2} are similar in magnitude.

\begin{figure}
        \centering
        \includegraphics[width=\linewidth]{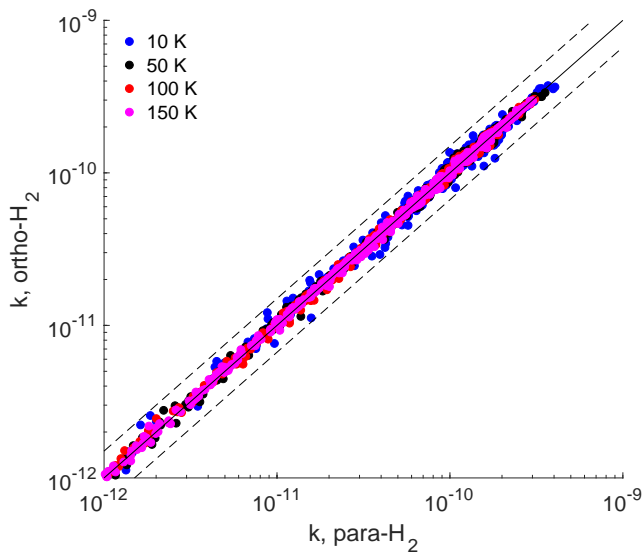}
        \caption{Comparison between rate coefficients (in units of cm$^3$~s$^{-1}$) for the collisional excitation of \ch{CF+} by para- and ortho-\ch{H2}. Rotational transitions involving the first 22 rotational levels of \ch{CF+} are presented at four different temperatures. The two dashed lines delimit the region where the rate coefficients differ by less than a factor of 1.5.}
        \label{paraortho}
\end{figure}

This similarity between para- and ortho-\ch{H2} rate coefficients is explored further in Figure \ref{paraortho}, which compares para- and ortho-\ch{H2} collisional rate coefficients for the rotational excitation of \ch{CF+} at four different temperatures (10, 50, 100, and 150~K), and for all transitions involving the first 22 rotational levels of \ch{CF+}.
As highlighted by the two dashed lines, differences between the two sets of data are quite small and do not exceed 50\%.
The agreement improves with increasing temperature, with differences of less than 10\% being observed for temperatures above 50~K.
It should also be noted that for the most dominant transitions ($k>2\times10^{10}$ cm$^3$~s$^{-1}$), differences stay lower than 10\% for all the temperatures explored.
This global agreement between para- and ortho-\ch{H2} rate coefficients confirms what has already been observed for a wide variety of ion--molecule collisions  \citep{masso:2014, walker:2017, denis-alpizar:2020, balanca:2020,klos:2011}.

\begin{figure}
        \centering
        \includegraphics[width=\linewidth]{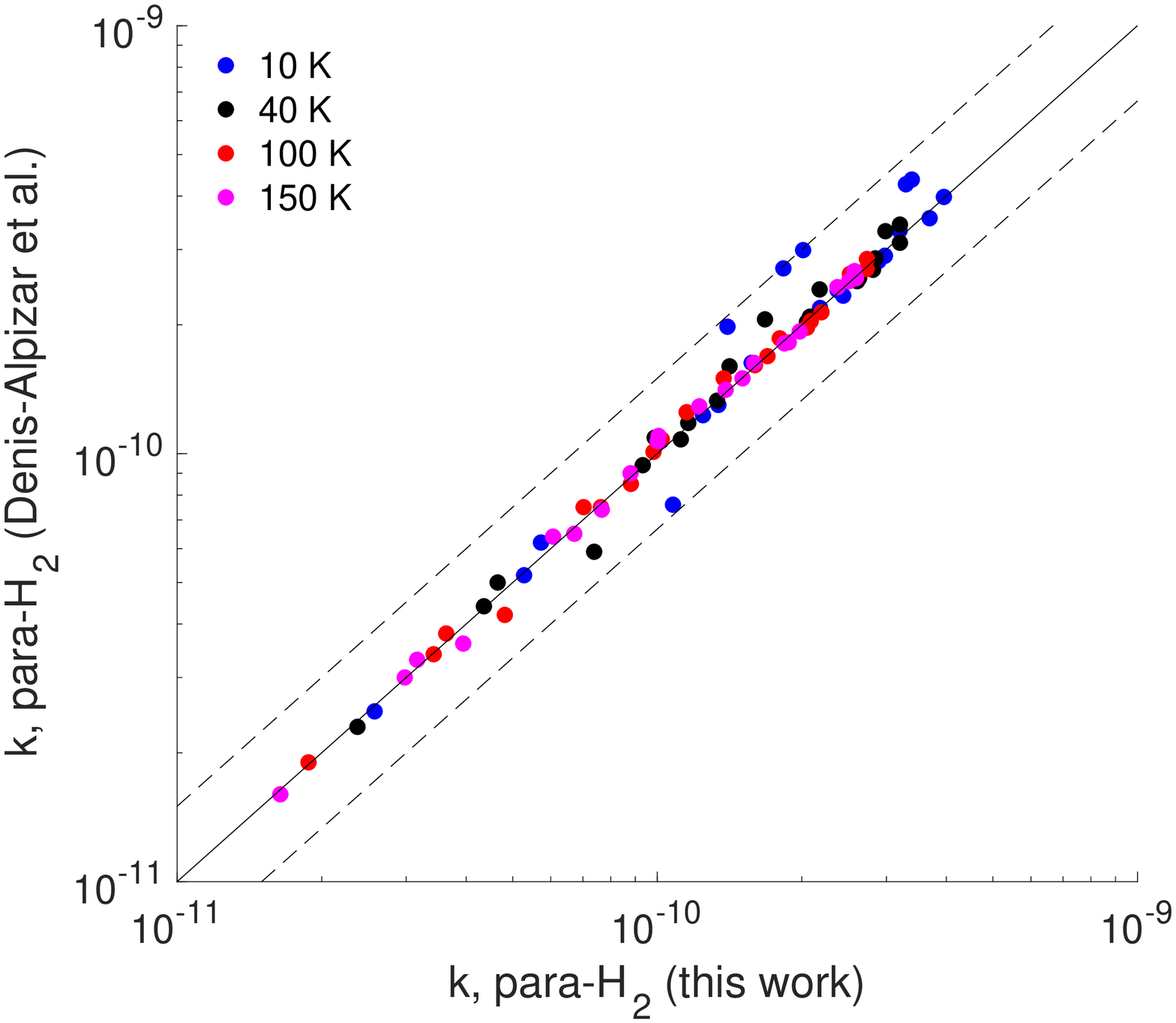}
        \caption{Comparison between rate coefficients (in units of cm$^3$~s$^{-1}$) for the collisional deexcitation of \ch{CF+} by para-\ch{H2} obtained in this work and those obtained by \citet{denis-alpizar:2019}. Deexcitation transitions involving the first seven rotational levels of \ch{CF+} are presented at four different temperatures. The two dashed lines delimit the region where the rate coefficients differ by less than a factor of 1.5.}
        \label{paraDA}
\end{figure}

In Figure \ref{paraDA}, we present a comparison between rate coefficients for the collisional deexcitation of \ch{CF+} by para-\ch{H2} obtained in this work and those obtained by \citet{denis-alpizar:2019}.
The agreement between the two sets of data is good overall, with differences of less than 10\% generally observed at high temperatures ($\ge$40~K).
At low temperatures, and, in particular, for the most dominant transitions, differences up to 50\% can be seen.
These differences can likely be attributed to the different levels of electronic structure calculations used to generate the PESs, as well the use of an averaged version of this PES, neglecting the \ch{H2} structure. Therefore, we recommend the use of our new rate coefficients in astrophysical applications.

\section{Excitation and radiative transfer study}

\begin{figure*}
        \centering
        \includegraphics[width=.75\linewidth]{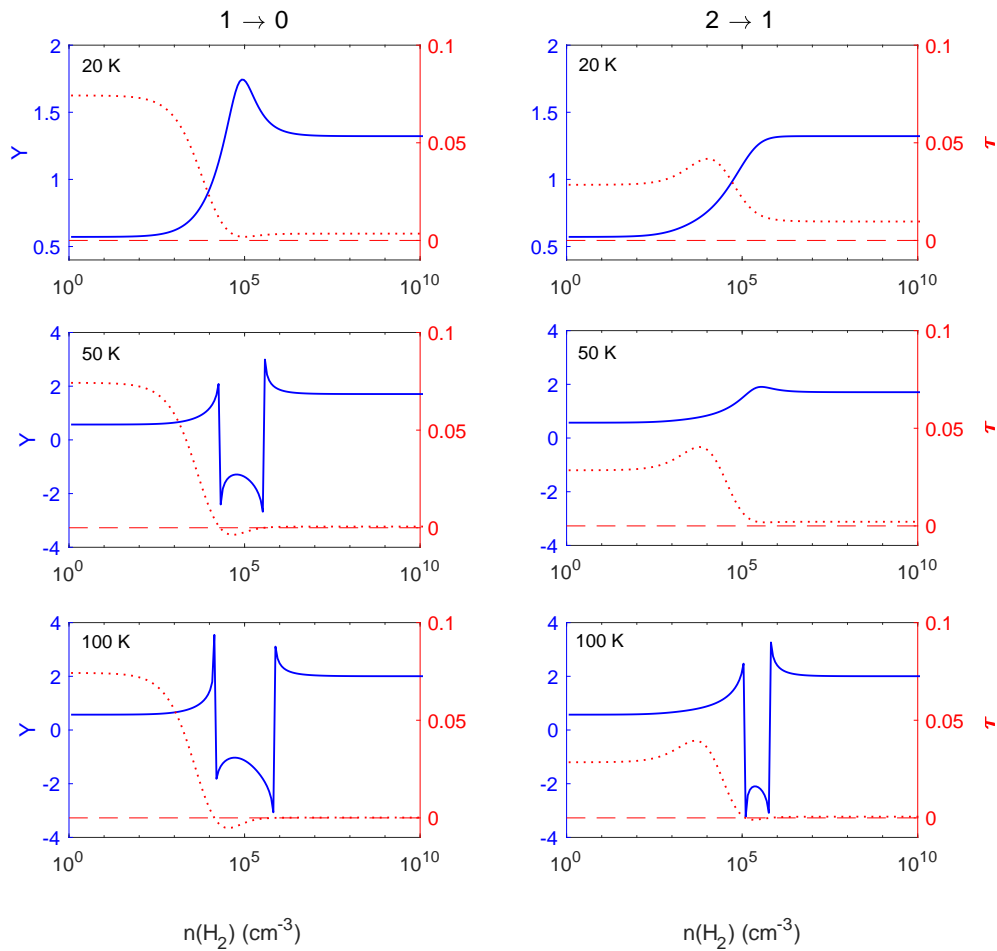}
        \caption{Variation of excitation temperature (blue solid lines) and optical depth (red dotted lines) as a function of \ch{H2}-density for the $1\rightarrow0$ (left panels) and $2\rightarrow1$ (right panels) \ch{CF+} rotational transitions at kinetic temperatures of 20~K (upper panels), 50~K (middle panels), and 100~K (lower panels), and a \ch{CF+} column density of $N($\ch{CF+}$) = 1\times10^{12}$~cm$^{-2}$.
                In order to better display the large amplitude of its variation, the excitation temperature is represented on this figure as $Y=\frac{T_{\text{ex}}}{\left| T_{\text{ex}} \right|} \times \text{log}_{10}\left( 1 + \left| T_{\text{ex}} \right| \right)$.}
        \label{laet}
\end{figure*}
\begin{table*}
        \centering
        \begin{threeparttable}
                \begin{tabular}{ccccccc}
                        \hline
                        &  & \multicolumn{3}{c}{Critical density\tnote{a}~\ (10$^5$ cm$^{-3}$)} & \multicolumn{2}{c}{$W$ (mK.km.s$^{-1}$)} \\ 
                        Rotational line& Frenquency (GHz) & 20~K & 50~K &100~K & Horsehead & Orion Bar \\ 
                        \hline 
                        $1 \rightarrow 0$ & 102.58748&  0.28 & 0.39 & 0.48 & 150$\pm$20 & 86$\pm$10 \\ 
                        $2 \rightarrow 1$ & 205.17445&  1.34 & 1.78 & 2.46 & 290$\pm$40 & 337$\pm$13 \\ 
                        $3 \rightarrow 2$ & 307.7443  & 5.16 & 5.80 & 7.04 & - & 428$\pm$34\\
                        $5 \rightarrow 4$ & 512.8465   & 11.76 & 13.36 & 15.07 & - & 100$\pm$20
                \end{tabular} 
                \caption{Observed \ch{CF+} integrated line intensities in the Horsehead and Orion Bar PDRs \citep{neufeld:2006,guzmanIRAM30mLineSurvey2012,nagy:2017}.}
                \label{ws}
                \begin{tablenotes}
                        \item [a] Critical density: $n_{\rm cr} = \frac{A_{ij}}{k_{ij}}$, where $i$ and $j,$ respectively, designate the initial and final rotational number, $A$ is the Einstein coefficient, and $k$ is the collisional rate coefficient.
                \end{tablenotes}
        \end{threeparttable}
\end{table*}

\begin{figure*}
        \centering
        \includegraphics[width=.49\linewidth]{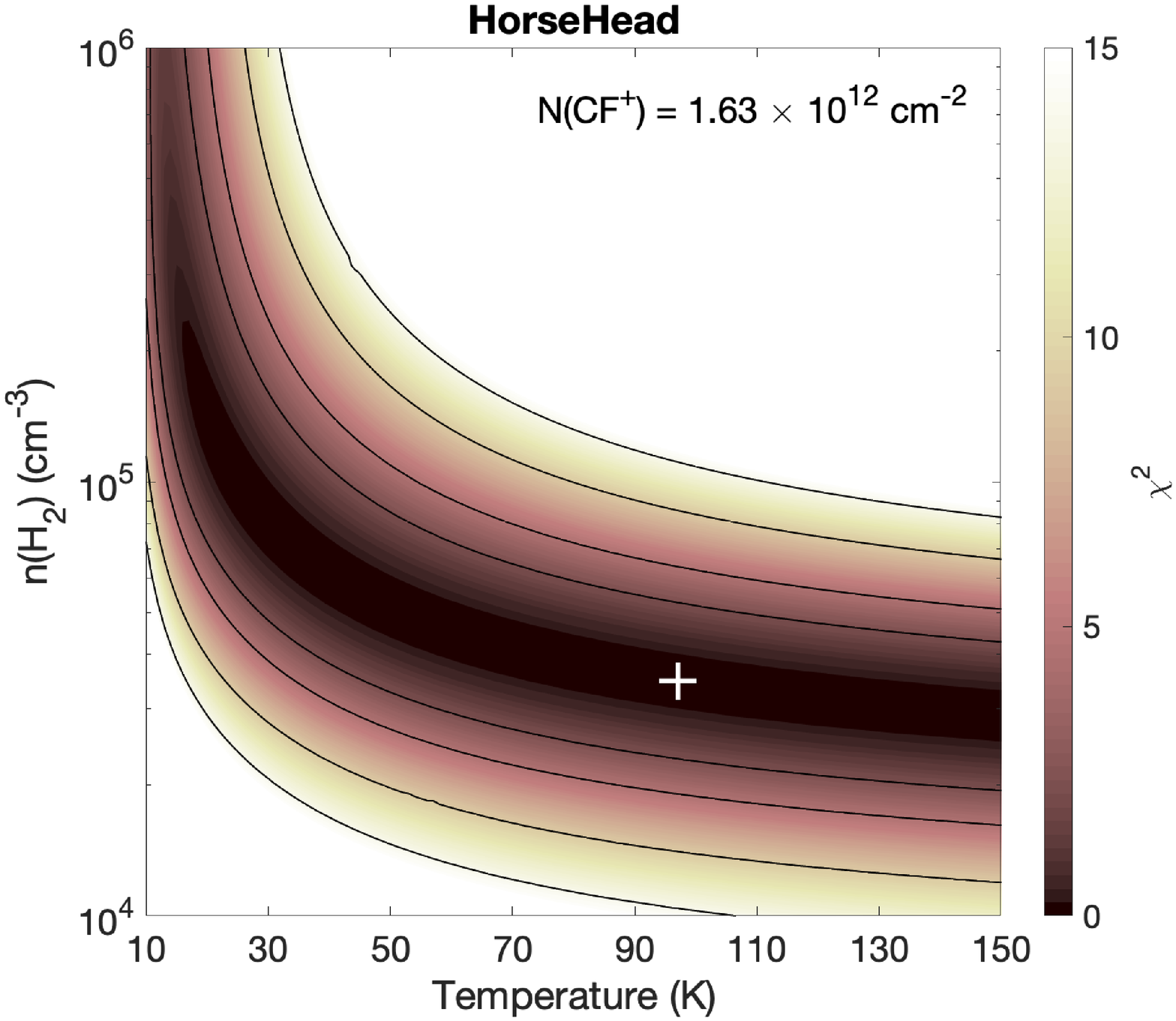}
        \includegraphics[width=.49\linewidth]{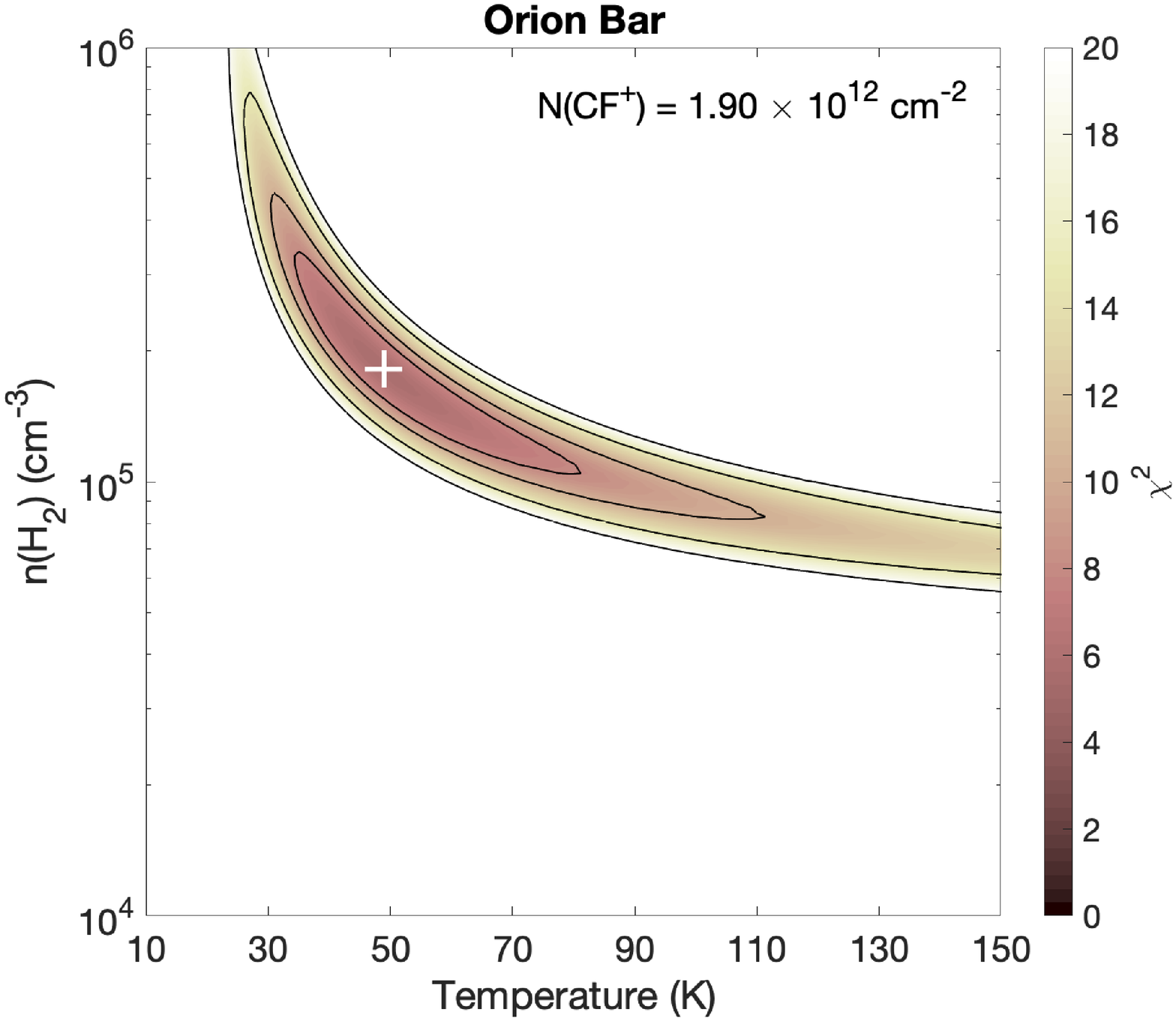}
        \caption{$\chi^2$-value as a function of the \ch{H2} column density and gas kinetic temperature for the Horsehead (left panel) and Orion Bar (right panel) PDRs. The solid black lines represent confidence contour levels of 63.3\%, 90.0\%, 99.0\% and 99.9\%. The sets of parameters ($T$, $n$(\ch{H2})) which give minimum $\chi^2$ are indicated with a white + symbol.}
        \label{radexfit}
\end{figure*}
In order to test the impact of the new collisional rate coefficients in astrophysical applications, the \textsc{radex} \citep{vandertak:2007} code was used to perform non-LTE radiative transfer calculations\footnote{The collisional rate coefficients are made available as a \textsc{radex} datafile in the LAMDA database \citep{2005A&A...432..369S}.} using the escape probability formalism, assuming an isothermal and homogeneous medium.

The radiation field value was taken as the cosmic microwave background at 2.73~K, and the line width was set at 1 km/s.
Energy levels, transition frequencies, and Einstein $A$ coefficients were taken from the Cologne Database for Molecular Spectroscopy (CDMS; \cite{muller:2005}).

Collisional deexcitation rate coefficients for the first 22 rotational levels of \ch{CF+} (231 collisional transitions) obtained in this work were included for temperatures up to 150~K.
Note that because of the absence of nuclear spin-dependence of the rate coefficients, we only considered collisional excitation of \ch{CF+} through para-\ch{H2} collisions in the calculations.

It should also be noted that electron collisions were not included in these radiative transfer calculations.
To the best of our knowledge, there is no available data for the collisional excitation of \ch{CF+} by electrons.
Nevertheless, while \ch{CF+} is detected in regions of high ionization fractions, with the electron density given by \mbox{$n_e = n_H \times X(\ch{C+}) \sim 2n(\ch{H2})\times10^{-4}$ cm$^{-3}$}, as for many molecular ions, \ch{CF+}--\ch{H2} collisional rate coefficients are particularly large: \mbox{$>10^{-10}$ cm$^3$~s$^{-1}$}.
This implies that in order for electronic collisions to contribute to the excitation process of \ch{CF+} molecules, electron rate coefficients should be greater than a few $10^{-6}$ cm$^3$~s$^{-1}$, and even greater than \mbox{$10^{-5}$ cm$^3$~s$^{-1}$} to dominate.

The \ch{H2}-density variation of the excitation temperatures and population levels were computed for a grid of kinetic temperatures ranging from 10 to 150~K and column densities ranging from $10^7$ to $10^{15}$~cm$^{-2}$.
In Figure \ref{laet}, we represent the \ch{H2}-density variation of the excitation temperature\footnote{\mbox{$n_u/n_l=\,g_u/g_l\,exp[-E_{ul}/k_{\rm B}\,T_{\rm ex}]$}, where $n_u$ ($n_l$) represents the population of the upper (lower) level, $g_u$ ($g_l$) the degeneracy of the upper (lower) level, $E_{ul}$ the difference of energy between the upper and lower levels, and $k_{\rm B}$ the Boltzmann constant. $T_{\rm ex}=T_{\rm k}$ corresponds to LTE conditions.} and the optical depth $\tau$ for the $1\rightarrow0$ and $2\rightarrow1$ \ch{CF+} rotational lines.
We explore kinetic temperatures of 20, 50, and 100~K, and a \ch{CF+} column density of $10^{12}$~cm$^{-2}$, corresponding to the typical value inferred from previous observations \citep{neufeld:2006,guzmanIRAM30mLineSurvey2012,nagy:2017} in PDRs.

At low temperatures ($T_{\rm k}=20$~K), the  $1\rightarrow0$ rotational line exhibits supra-thermal emission ($T_{\rm ex}>T_{\rm k}$) in the \ch{H2}-density range of $\sim 10^4-10^6$~cm$^{-3}$.
At higher kinetic temperatures (50 and 100~K), the excitation temperature becomes negative in approximately the same \ch{H2}-density range.
The same behavior is observed for the $2\rightarrow1$ line in a narrower \ch{H2}-density range and at higher temperatures:
supra-thermal excitation appears at $\sim$~50~K in the \ch{H2}-density range of $\sim 10^5-10^6$~cm$^{-3,}$ while negative excitation temperatures are observed at a kinetic temperature of 100~K.
This negative excitation temperature behavior indicates a level population inversion ($n_u/g_u>\,n_l/g_l$).
However, as can be seen in Figure \ref{laet}, at the relatively low \ch{CF+} column densities present in prototypical PDRs, the opacities of the inverted lines remain very low ($-\tau << 1$) for all the explored physical conditions.
This implies a negligible maser amplification insufficient to exhibit an observable effect.

Finally, we used these \textsc{radex} models to infer the H$_2$ gas density at the UV-illuminated surfaces of the Horsehead and Orion Bar PDRs from the observed \ch{CF+} line emission.
In particular, we tried to fit the \ch{CF+} emission spectra reported by \citet{neufeld:2006}, \citet{guzmanIRAM30mLineSurvey2012}, \citet{guzman:2012a}, and \cite{nagy:2017} summarized in Table \ref{ws}.

In order to determine the set of parameters ($T$, $n$(\ch{H2}) and $N$(\ch{CF+})) that best reproduces the observations, we calculated the $\chi^2$-value for each set of parameters as follows:
\begin{equation}
\chi^2 = \sum_{i=1}^n \left( \frac{W_i^{\rm obs} - W_i^{\rm calc}}{\sigma_i} \right)^2 
,\end{equation}
where $n$ is the number of observed \ch{CF+} rotational lines, $W^{\rm calc}$ is the integrated line intensity obtained from \textsc{radex} simulations, $W^{\rm obs}$ is the observed integrated line intensity reported in the papers, and $\sigma$ is the uncertainty on the observed value.
The best fit is then obtained by minimizing the $\chi^2$ values.

These fits implicitly assume that the observed \ch{CF+} emission is spatially extended (i.e., it fills the beam of the telescope at each observed frequency).
This is a reasonable assumption because  \ch{CF+} is expected to arise from the extended \ch{C+} layers at the edge of PDRs  (see models of Neufeld \& Wolfire 2009, Guzm\'an et al. 2012b).

For both sources, the best-fit value for the \ch{CF+} column density $N$(\ch{CF+}) allowing us to reproduce the observed line strengths were $1.63\times 10^{12}$ cm$^{-2}$ and $1.90 \times 10^{12}$ cm$^{-2}$ for the Horsehead and Orion Bar regions, respectively.
In Figure \ref{radexfit}, we present the $\chi^2$-value as a function of the \ch{H2} gas density and gas kinetic temperature for both Horsehead and Orion Bar PDRs at the corresponding \ch{CF+} column densities.
In this Figure, we also represent contour levels where the probability of enclosing the correct parameters $T$, $n$(\ch{H2}) and $N$(\ch{CF+}) is 63.3\%, 90.0\%, 99.0\%, and 99.9\%.
This is achieved by considering the regions where $\chi^2  \le \chi^2_{\rm min} + \Delta \chi^2$ for $\Delta \chi^2$ values of 2.3, 4.6, 9.2, and 13.8 \citep{1976ApJ...208..177L}.
The sets of parameters ($T$, $n$(\ch{H2})) that give minimum $\chi^2$ (i.e., the best fit) are indicated with a white + symbol:
we obtain kinetic temperatures of 100~K and 50~K, and \ch{H2} gas densities of $3.5\times10^{4}$~cm$^{-3}$ and $1.8\times10^{5}$~cm$^{-3}$, for the Horsehead and Orion Bar PDRs, respectively.

In those two prototypical PDRs, the gas temperature of the \mbox{\ch{C+}-emitting} layers (likely similar to the  \ch{CF+}-emitting layers at the UV-illuminated cloud surface) are observationally well constrained to $T_{\rm k}=60-100$~K  in the Horsehead  \citep{Pabst:2017} and $T_{\rm k}=150-200$~K in the Orion Bar \citep{Goicoechea:2016}. 
We note that a more refined determination of the gas density would require one to spatially resolve the true size of the \ch{CF+} -emitting regions (e.g., with ALMA) and carry out PDR-depth dependent excitation models. 
If we assume that the \ch{CF+} emission in the Orion Bar arises from a \mbox{10’’-width} filament (similar to other molecular ions such as \ch{SH+}, Goicoechea et al. 2017) and correct the observed line intensities by the appropriate beam filling factors, we obtain reasonable fits consistent with gas temperatures of about 100~K and \ch{H2} gas density n(\ch{H2})$\sim7\times 10^4$ cm$^{-3}$.

\section{Summary}
In summary, we computed rate coefficients for inelastic collisions of \ch{CF+} by both para- and ortho-\ch{H2} for temperatures up to 150~K and rotational transitions between the first 22 levels of the \ch{CF+} molecule ($0\le j_1 \le 21$).
We used these new rate coefficients in non-LTE excitation and radiative transfer calculations that reveal inversion population and weak maser \ch{CF+} emission at gas physical conditions corresponding to those in regions where this molecule can be observed.
Finally, we were able to constrain gas density at the surface of the Horsehead and Orion Bar PDRs by reproducing their observed \ch{CF+} line intensities.

\begin{acknowledgements}
        F.L. acknowledges financial support from the Institut Universitaire de France. We acknowledge the Programme National Physique et Chimie du Milieu Interstellaire (PCMI) of CNRS/INSU with INC/INP co-funded by CEA and CNES. 
This work was granted access to the Occigen HPC resources of CINES under the allocation 2019 [A0070411036] made by GENCI. 
J.R.G. thanks the Spanish MICIU for funding support under grants \mbox{AYA2017-85111-P} and \mbox{PID2019-106110GB-I00}.
R.D. is supported by the US Department of Energy Office of Science, Office of Basic Energy Sciences (Award DE-SC0019740).
\end{acknowledgements}

\bibliography{cfph2.bib}
\bibliographystyle{aa}
\end{document}